# Expectations and Reality: Why an enterprise software system didn't work as planned

**David Greenwood and Ian Sommerville[1]**

**Abstract**  Over two decades, we and other research groups have found that ethnographic and social analyses of work settings can provide insights useful to the process of system analysis and design. Despite this, ethnographic and social analyses have not been widely assimilated into industry practice. Practitioners tend to address sociotechnical factors in an ad-hoc manner, often post-implementation, once system use or outcome has become problematic. In response to this, we have developed a lightweight qualitative approach to provide insights to ameliorate problematic system deployments. Unlike typical ethnographies and social analyses of *work activity* that inform systems *analysis* and *design*; we argue that analysis of *intentional* and *structural factors* to inform *system deployment* and *integration* can have a shorter time duration and yet can provide actionable insights.  We evaluate our approach using a case study of a problematic enterprise document management system within a multinational systems engineering organization. Our findings are of academic and practical significance as our approach demonstrates that structural-intentional analysis scales to enable the timely analysis of large-scale system deployments.

## 1 Introduction

There is recognition within the sociotechnical systems community that practitioners' needs are not being met by sociotechnical design methods [1]. Practical issues such as lack of industrial take up and the recognition of theoretical issues such as the 'design fallacy' motivate this. The design fallacy is the presumption that the primary solution to meeting users' needs is to develop ever more sophisticated social analyses to inform design [2]. Studies reveal that sociotechnical integration is not the *passive* 'diffusion' of a system into an organization, but is strongly dependent on the *active* adaptation and reconfiguration of the technical system and work practices during and after *deployment* [3,4]. This aspect of sociotechnical systems engineering is largely unexplored from an engineering perspective and, as a result, we (as a community) are devoid of timely scalable engineering techniques to pinpoint sociotechnical problems post-deployment, or indeed techniques to in-

David Greenwood
School of Computer Science, University of St Andrews, UK, email: dsg22@cs.st-andrews.ac.uk
Ian Sommerville
School of Computer Science, University of St Andrews, UK, email: ifs@cs.st-andrews.ac.uk

form the design of deployment strategies to ensure that adaptation and reconfiguration can occur in a particular context.

In response to this, we have developed an interview-based approach to pinpoint sociotechnical problems post-deployment. Our approach elicits and analyses the interactions between structural and intentional elements present in a situation that mediate stakeholders' use(s) of a technology to facilitate their work. Our approach contrasts with ethnographies and social analyses that typically inform system development because it does not involve detailed observation or analysis of tasks being performed. Our fieldwork demonstrates that structural-intentional data is: i) sufficient to diagnose problematic interactions between a system, intentional elements (roles, responsibilities, beliefs including those about the deployment) and structural elements (community norms, division of labour) in a situation; ii) sufficient to suggest practical interventions to ameliorate the situation e.g. the creation of networks of champions to reinforce structural-intentional elements and thus support system adoption.

## 2 Social Analysis to Inform Systems Development

Our experiences of social analysis started in the early 1990s with ethnographic studies of air traffic control [5] and later studies of financial institutions [6]. Subsequently our research focused on the development of a presentation framework to make ethnographic findings more accessible to systems analysts and designers. This work evolved into a structured approach call COHERENCE to enable non-ethnographers (typically systems analysts and designers) to organize their observations and develop use-cases to inform systems analysis and design [7,8].

In parallel to our efforts, other social analyses for informing systems development were developed e.g. ETHICS [9], MULTIVIEW [10] and the I* FRAMEWORK [11] . These approaches all shared the overarching aim of incorporating social analysis into systems analysis and design. Whilst each of these methods has had some laudable success they have not been widely assimilated by industry [12]. Our experiences suggest that practitioners address sociotechnical factors in an ad-hoc manner, often post-deployment, once system use or outcome becomes problematic [13]. Approaches that primarily inform analysis and design have never met this need.

We believe that a more fruitful approach may be to develop social analyses to inform the *deployment* and *integration* phase of systems development because: i) it fits with the ongoing trend for COTS where organisations change to fit a system; ii) studies of organisational change suggest that the manner by which a change is implemented has a significant effect on its completion rate and sustained adoption rate [14]; iii) the deployment process represents a significant sociotechnical engineering challenge that is likely to benefit from being informed by social analysis as it involves the creation and configuration of sociotechnical structures such as networks of champions, steering committees, wikis and so on [3,4].

Although the I* FRAMEWORK does provide some capability to analyse problematic deployments it is undesirable to do so because of a number of practical limitations: i) its task/activity centric approach requires in-depth study of work activity which is precisely the kind of time intensive analysis that makes social analysis impractical in many system developments; ii) its task/activity centric approach abstracts away structural elements that research suggests is important to deployment [15]; iii) its task/activity centric approach, as of yet, does not scale well to support the analysis of large-scale systems [16].

Instead we advocate a structural-intentional approach as it offers a different set of trade-offs that will be discussed in section 5. We justify our switch from a task/activity centric view to a structural-intentional view as qualitative and quantitative studies of technology acceptance, use and appropriation demonstrate a strong relationship between technology use and structural-intentional elements [17-19,4].

## 3 Research Design

A case study approach was selected because our aim was to test the hypothesis that 'structural-intentional analysis enables the timely analysis of large-scale deployments by identifying incompatibilities between structural and intentional elements'. A case study approach was deemed appropriate since our aim was to test our hypothesis in a real-world corporation with a problematic system. We collected our data using 16 hour-long semi-structured interviews of the system's stakeholders comprising open-ended and closed questions. We interpreted our data using a variant of cultural-historical activity theory called the 'activity space' [20,21]. We chose this framework as it gives primacy to the interrelationships between intentional and structural aspects of a situation.

### 3.1 The 'Activity Space' Framework

The 'activity space' framework [20,21] is a framework for structuring data and identifying problematic intentional and structural aspects of a system. The framework comprises three intentional constructs comprising: mediators (tools, beliefs, skills); subjects (roles, responsibilities) and object(ive)s. And three structural constructs comprising: rules (formal / informal norms); community (actors involved in a situation); and the division of labour (how work is divided). According to 'activity space' theory the outcome of a situation (e.g. a deployment) is brought about by interactions between actors behaviour(s). Each actor's behaviour is mediated by intentional and structural elements. So problematic situations can arise when tensions exist within and between actors' intentional and structural elements. By understanding these tensions a situation can be modified to change the outcome.

## 3.2 The Organization and the Sociotechnical System

The fieldwork was performed at three different sites of a multinational system engineering corporation that we will call 'Company A'. Their main work activity comprises the design, manufacture and maintenance of specialist electro-optical components and systems. The organisation is divided into a number of functional groups that come together under a project structure to produce customer deliverables e.g. components, systems and documents. The design of components and systems is a collaborative activity and the sharing of documents is considered to be an important aspect of this activity by those involved.

'Company A' deployed an electronic document management (EDM) system in the early 2000s as it was perceived by the IT director that an EDM system would be more advantageous than using shared folders on a file server to exchange documents. There was a perception that the introduction of the system would bring about greater visibility and awareness of work rather than having different teams and functions working in information silos. Within projects it was envisioned that EDM would be an up-to-date repository of all project documentation. Teams would store their documents in personal working areas and upload them to standardised locations in standardised EDM project file structures.

When we visited the organisation in 2010 the EDM was perceived by engineering management to be problematic due to "sociotechnical factors". The use of the system was mandatory so all projects had an EDM project area but the extent that documents were being uploaded from working areas to the EDM project areas varied between teams. In addition to this the use of the EDM file structure varied between teams, as did the location of files within the file structure. As our investigation unfolded it became clear that engineering management perceived the system to be problematic because teams did not use it in a "common way".

## 3.3 Data Collection and Analysis

We collected our data using 16 one-hour semi-structured interviews of the document management system's stakeholders. Interview participants were selected on the basis of availability by a facilitator within the organisation. The interviews comprised a set of open-ended questions and a set of closed questions comprising 7 point semantic differential scales and 7 point Likert scales. A copy of our survey can be found here (http://www.cs.st-andrews.ac.uk/~dsg22/P/EDM_Survey.pdf).

Interviews were digitally recorded and transcribed when permitted. The open-ended interview questions were designed to elicit the relationship between the participant's view of the their work (role, responsibilities, their day-to-day activities, most serious work challenges) and the deployed system (their history with the system, which responsibilities/activities the system helps them accomplish, how it does so, what problems it introduces to their work, how the system impedes their responsibilities and activities). The closed interview questions elicited the relationship between the participant and the system by exploring aspects of IT systems that are associated with intention to use (performance expectancy, effort expectancy, information quality, system quality, support quality, system usage policy) and aspects of organizational change that can lead to conflict (interfering with roles, goals, values, resources, capabilities/skills, job satisfaction, status, procedural justice, distributive justice, importance, ownership).

Dialogue mapping was then used to organize interview data into more abstract units of information. Dialogue maps were compared to the participant's responses to closed questions to corroborate findings. Dialogue maps were compared across participants to identify themes. The 'activity space' framework was then used to structure the findings and provide a lens for identifying tensions between different elements within the situation.

# 4 Our Findings

We report our findings in three parts: the first part identifies *software* usability issues, such as UI issues, that frustrate the use of EDM regardless of the software's specific configuration; the second part identifies *system* usability issues to do with the deployed configuration fitting the existing work environment; the third part reports on the *structural-intentional* issues that frustrate the use of EDM. In the third part, EDM is viewed as a resource that mediates (enables / constrains / transforms) work activity. In contrast to the first and second parts, the issues raised will highlight underlying tensions that result in issues or challenges that impede the use of EDM in a "common way" as desired by engineering management.

## *4.1 Software Usability Issues*

We found that the following aspects confounded the usability of the tool in both experienced and novice/infrequent users. The consequences of these issues were typically frustration and/or perceptions of wasted time.

**Table 1.** Aspects Detrimental to Usability According to Experienced Users

| # | Aspects detrimental to usability according to experienced users |
|---|---|
| 1 | Requires a login separate from workstation login. |
| 2 | Web-based interface is slow to respond to user interaction. |
|   | -Screen updates and file uploads are perceived to be slow or freeze. |
| 3 | Files can only be uploaded individually using the web-based interface. |
| 4 | Files are rendered poorly when viewed in using the web-based interface. |

| 5 | Search feature does not return expected results. |
|---|---|
| 6 | Web-based interface has screen-rendering issues when used with browsers other than Internet Explorer 6. |

**Table 2.** Additional aspects detrimental to usability according to new / infrequent users

| # | Additional aspects detrimental to usability according to new / infrequent users |
|---|---|
| 1 | Menus are cluttered and there is no obvious feature prioritisation to guide novice/infrequent users. |
| 2 | Search query presentation is difficult to understand -E.g. Use of MIME types. |
| 3 | The 'look & feel' of the web-based interface is dissimilar to the 'drag & drop' interfaces that end-users are generally accustomed to. |

End-users found that the above issues to be slightly problematic but in general they perceive them to have a minor effect on their overall productivity, job satisfaction, speed of accomplishing work activity, and effort to use EDM.

- Most participants surveyed agreed or strongly agreed that EDM takes little effort to use on their part. They reported that EDM did not *significantly* improve nor worsen their individual productivity, responses were mainly distributed around no effect, or slight positive or negative effects.
- Most participants agreed that EDM did not *significantly* slow down or speed up their speed of accomplishing activities. Responses were distributed equally between no difference, slower and faster.
- Most participants reported that EDM neither favourably nor adversely affects their job satisfaction. However participants did report that the user interface does not meet their needs and is slightly problematic. And that the search facility does not meet their needs and is slightly problematic.

These mixed responses indicate that although the system has a number of frustrating and/or timing wasting usability issues, the majority of users we interviewed found that it did not significantly interfere with either their overall productivity or job satisfaction. These findings are perhaps surprising as management perceive the system to be problematic. This difference is explained by the fact that the extent that each team uses EDM is in accordance with their own approach and therefore they use the system in a manner that is acceptable to them (as a team) but not necessarily in a manner that is desired by management.

## 4.2 System Usability Issues

We found that the following aspects confounded the usability of the tool in both experienced and novice/infrequent users.

**Table 3. Aspects detrimental to system usability**

| # | Aspects detrimental to system usability |
|---|---|
| 1 | EDM is perceived to be more time consuming to use for storing documents in comparison to shared drives, or personal areas, due to the software usability issues identified. |
| 2 | EDM has been configured to offer standardised folder structures however users struggle to understand where to put their documents within these structures. They perceive that there are a variety of possible locations, which makes remembering and sharing the location of a document problematic. This interpretative flexibility enables the use of EDM in contrasting and inconsistent ways. |
| 3 | EDM project areas have no built-in document registers making it difficult to establish what documents are within a project area and which are missing. Lack of a document register is seen be problematic because of inconsistent use of the standardised folder structures which makes finding files on the basis on their expected folder location impractical. |
| 4 | EDM has practical limitations on the number of files in a single folder as this can cause freezing or degrade the performance of the search facilities. This has been mitigated on the most part by using a folder structure. |
| 5 | EDM runs on servers within a 'restricted' network and so cannot be accessed by all parts of the organization. In a number of situations this results in end-users having to use other IT resources for document sharing undermining the purpose of the EDM. |

Despite the issues identified above, the participants that we interviewed reported that, in general, the use of EDM does facilitate their working practices, they are supportive of continuing investment and development of EDM, and that EDM is considered to be slightly important, or important, to their interests and responsibilities. This indicates that despite the systems shortcomings it was recognised by those that we interviewed as a valuable tool that supports work. Again these findings may be surprisingly considering that engineering management perceive the system to be problematic. However this again highlights that end-users use the system in a manner that is acceptable to them but not necessarily in a "common way" as desired by engineering management.

## 4.3 Structural-intentional Issues

Overall we found that the extent and nature of EDM use varies on a project-by-project basis and that the nature of use is dependent upon individual programme managers and engineering teams. Engineering management perceive this to be problematic and at the time were pursuing an improvement strategy that encourages standardization of EDM use. We analysed this problematic situation as the outcome of the following interacting elements: roles; objectives; mediators; division of labour; communities; and rules.

**Roles**

- There is a potential for conflict between the roles of engineering management and programme management. Whilst it is the responsibility of engineering management to take a strategic (long-term) view and run improvement projects. This is at tension with the tactical (shorter-term) responsibilities of programme management. Introducing change, even when successful, can cause short-term productivity degradation as changes are being 'bedded in'. This can be at odds with programme managers' contractual obligations such as milestones.

**Objectives**

- The objectives of engineering management, programme management and engineers are aligned such that their overall objectives are positively dependent. This means it is in all parties interests to coordinate their activities as one parties success contributes to the success of the other parties.
- Whilst objectives are positively dependent there is however scope for process conflict and our study suggests that it may be occurring. For example engineer managements' objective of improving deliver time is compatible with programme managements' objective of meeting contractual obligations however the way in which the objective is pursued, such as modifying the EDM may interfere with meeting a contractual obligations if not carefully coordinated.

**Mediators**

- EDM is acknowledged by all communities to suffer from usability issues at both the software and system level that ultimately results in frustration.
- These usability issues provided a motive in some communities to use a shared drive rather than EDM.

**Division of labour**

- A matrix structure is a conduit of tensions. Engineers are within the focal point of matrix and so they resist change when incompatible demands are placed.
- The division of labour can result in disconnects of responsibility or ownership. This has occurred with respect to the domestication of EDM.

**Communities**

- Communities have emerged around the division of labour and thus roles/responsibilities e.g. program management, engineers. There is a divergence between the viewpoints of each of these communities with respect to the value of EDM and the salience of its capabilities and purpose.

**Rules**
- There is a strong practice culture rather than a process culture within programme management and engineering. This means that work is performed on the basis of norms (e.g. individual and shared experience of what has happened in the past) rather than following explicit 'rules' (e.g. referring to process documentation).

We found that these tensions interacted to create four vicious circles that are contributing to sustaining the problematic situation [22].

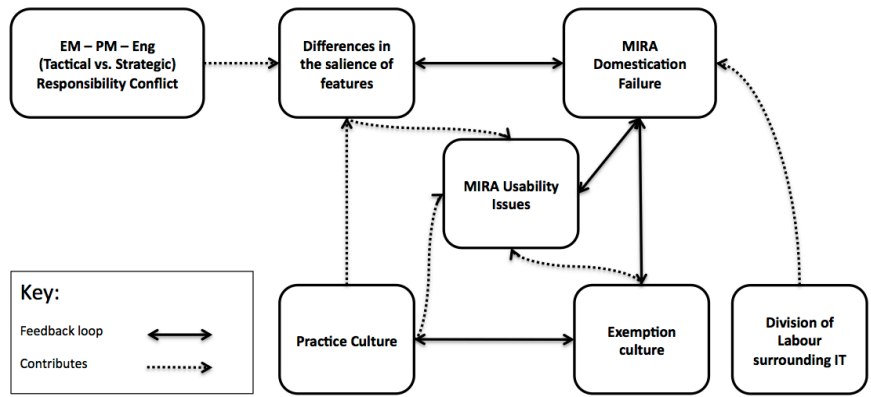

**Fig. 1** - Interactions of the socio-technical issues identified

The first vicious circle occurs between the differences in saliencies of usability issues and the domestication of EDM. The programme managers and engineers do not value the visibility and control features of EDM as much as engineering management. This has resulted in the continued use of shared drives, rather than the use of EDM as intended, resulting in an absence of familiarity, adaptation and acceptance of EDM. Conversely as domestication has not occurred, programme managers and engineers did not have the opportunity to become familiar with the benefits and drawbacks of the visibility and control features.

The second vicious cycle occurs between domestication and usability issues. Because domestication did not occur users experience usability issues due to lack of familiarity or because of lack of adaptation to the tool over time. Conversely, users experience usability issues because of lack of domestication.

The third vicious circle occurs between the exemption culture and domestication. The culture of allowing projects to decide on the extent and nature of EDM use (exemption culture) has resulted in a lack of familiarity with the full capabilities of EDM. Consequently EDM is not perceived as an acceptable substitute to shared drives. Conversely because EDM has not been domesticated this has reinforced the culture of exemptions by permitting users not to use the tool.

The fourth vicious circle occurs between the practice culture and the exemption culture. As work is performed on the basis of norms (e.g. individual and shared experience of what has happened in the past) rather than explicit rule adherence it has made it acceptable for projects to exempt themselves from standard ways of working such as EDM. Conversely, since projects are permitted to exempt themselves from standard practices this reinforces the 'practice culture' as enacting an exemption is in itself an exercise of the primacy of experience over standards.

# 5 Discussion

Our study was designed to evaluate the following hypothesis: 'structural-intentional analysis enables the timely analysis of large-scale system deployments'. Our study supports this hypothesis as firstly we identified a number of tensions between structural and intentional elements and their interactions. Secondly we were able to make specific recommendations to ameliorate the deployment.

The recommendations we made comprised a six-step plan to address the vicious circles identified. Each individual recommendation can be described codified commonsense. E.g. to address the exemption culture we recommended identifying, implementing and enforcing a set of mandatory practices in coordination with programme management. E.g. to address lack of domestication we recommended assigning the responsibility to a high-level manager and creating institutional structures such as networks of champions and steering groups to guide familiarization and adaption of the work practices and the EDM. The value of the plan came from the fact that the *combination* of recommendations we made was tailored to the specific dynamics of the deployment environment.

Compared to a task/activity centric analysis approach we believe that structural-intentional analysis offers a number of useful tradeoffs. Firstly data collection can be of a shorter duration as a detailed understanding of tasks is not required and thus avoids time-consuming ethnography or process mapping. Secondly, the scale of the deployment under analysis can be much larger as data collection is rapid and data analysis can be supported through the use of off-the-shelf digraph visualisation and analysis tools that support large datasets. It is an open research question as to whether the task/activity centric approach of I* models can scale up to analyse large-scale systems [16].

The disadvantages of a structural-intentional approach, in contrast to a task/activity centric approach, is that it will not deliver insights with respect to the subtleties of task level interactions within a work environment e.g. distributed coordination, awareness, spatial and temporal organisation and so on. Nor does structural-intentional analysis enable modelling at the task and resource level so it does not represent how actors and resources are configured at a task level.

Despite these shortcomings, we believe that this trade-off is desirable as it makes structural-intentional analysis complementary to established task/activity centric analyses that inform information systems development. For example when time permits ethnography we expect structural-intentional analysis to provide complementary findings.

# 6 Conclusion

This work illustrates that structural-intentional analysis is a promising candidate as a scalable engineering technique for analysing and troubleshooting systems post-deployment. Our case study indicates that structural-intentional analysis has a number of attractive characteristics with respect to timeliness and scalability. Data collection appears more rapid than either process mapping or ethnography and data analysis appears extremely scalable as it can be supported through the use of off-the-shelf digraph visualisation and analysis tools. Our fieldwork demonstrates that structural-intentional data is: i) sufficient to diagnose problematic interactions between a system, intentional elements and structural elements; ii) sufficient to suggest practical interventions to ameliorate a deployment. We therefore advocate the structural-intentional approach as a candidate engineering approach for analysing and troubleshooting large-scale deployments.

Our conclusion is limited by the usual limitations of qualitative case study research. Case study research may not be generalisable and whilst every effort was taken to minimize investigator or participant bias, bias may be reflected in our findings.

There are many opportunities to further validate and develop the structural-intentional view of deployments. We encourage more case studies or action research to demonstrate its scalability and ability to ameliorate deployments in a variety of settings. We encourage comparative work between structural-intentional approaches and task/activity-centric approaches (such as I*) to explore their strengths and weaknesses. We also encourage the development of tools to support structural-intentional analysis.